\documentclass[aps,epsf,twocolumn,prl]{revtex4}
\newcommand{\bra}[1]{\left < #1\right|}
\newcommand{\ket}[1]{\left |#1\right >}

\newcommand{\opsandwich}[3]{\left < #1|#2|#3\right >}
\newcommand{\overlap}[2]{\left < #1|#2\right >}

\usepackage{amsmath}
\usepackage{amsfonts}
\usepackage{graphicx}
\usepackage{amsthm}
\usepackage{subfigure}
\usepackage{hyperref}
\usepackage[usenames]{color}

\usepackage{rotating}

\renewcommand{\v}[1]{\mathbf{#1}}

\newcommand{\lp}{\left ( }
\newcommand{\rp}{\right ) }
\newcommand{\lb}{\left [ }
\newcommand{\rb}{\right ] }

\newcommand{\hc}{\text{H.c.}}

\newcommand{\beq}{\begin{eqnarray*}}
\newcommand{\eeq}{\end{eqnarray*}}

\newcommand{\be}{\begin{eqnarray}}
\newcommand{\ee}{\end{eqnarray}}

\newcommand{\mc}{\mathcal}

\def\lsim{\mathrel{\rlap{\lower4pt\hbox{\hskip1pt$\sim$}}
    \raise1pt\hbox{$<$}}}                % less than or approx.

\def\gsim{\mathrel{\rlap{\lower4pt\hbox{\hskip1pt$\sim$}}
    \raise1pt\hbox{$>$}}}                % greater than or approx.

\begin{document}

\title{
On-site correlations in optical lattices: band mixing to coupled quantum Hall puddles}
\author{Kaden R.~A. Hazzard} \email{kh279@cornell.edu}
\affiliation{Laboratory of Atomic
and Solid State Physics, Cornell University, Ithaca, New York 14853}
\author{Erich J. Mueller}
\affiliation{Laboratory of Atomic
and Solid State Physics, Cornell University, Ithaca, New York 14853}

\begin{abstract}
 We extend the standard Bose-Hubbard model to
 capture arbitrarily
 strong on-site correlations.  In addition to being important for quantitatively modeling  experiments, for example, with Rubidium atoms, these correlations must be included  to describe more exotic situations.
 Two such examples are
 when the interactions are made large via a Feshbach resonance, or  when each site rotates rapidly, making a coupled array of quantum Hall puddles.   Remarkably, even the mean field approximation to our model includes all on-site correlations.  We describe how these on-site correlations manifest themselves in the
  system's
  global properties: 
   modifying the phase diagram and depleting the condensate.
   \end{abstract}


\maketitle

Optical lattice systems,  where a dilute atomic gas
is trapped in a periodic potential formed by interfering laser beams, provide a close connection between solid state systems and atomic physics~\cite{bloch:many-body-cold-atoms-review}.  The models used to describe these systems generally assume that each lattice site's wavefunction  is easily built up from
single particle states~\cite{jaksch:olatt}.  Here we argue that this approximation is inappropriate
to quantitatively model current experiments,
and sometimes fails
  more drastically, e.g., for resonant bosons.
 We show how to include arbitrary on-site correlations via a generalized Hubbard model, which can be approached by standard methods.
By construction, the mean field approximation
to our model
captures all on-site correlations.

Our method's key idea  is
to first
consider deep lattices, where lattice sites are isolated and then solve the few-body problem on each
site.  Next, truncating to this few-body problem's low energy manifold, we calculate how tunneling couples the few-body states on neighboring sites.
 The resulting theory resembles a Hubbard model, but with number-dependent hopping and interaction parameters.
  We show that
the corrections
to the ordinary Bose-Hubbard model
captured by this theory
are crucial to quantitatively describe current Rubidium experiments.  They become even more important when the 3D scattering length $a$ becomes a significant fraction of the size of the
Wannier states $\ell$, such as in recent experiments on Cesium atoms near a Feshbach resonance~\cite{chengchin:priv-comm}.
This approach is also essential to describe
more exotic on-site correlations; as one example,
one can  rotate each lattice site, creating a lattice of coupled ``quantum hall puddles"~\cite{popp:adiabatic-make-fqh,baur:fqh-pi-pulse}.
Related ideas 
can be applied to
double well lattices and coupled ``plaquettes" of four sites~\cite{barmettler:double-wells,jiang:double-well-decoherence-free}.
We explore the impact of the on-site physics on the extended system's phase diagram.

Our approach is most simply illustrated by a single-component Bose gas in a
cubic sinusoidal lattice potential $V_p(x,y,z) = V_0 \sum_{\eta=x,y,z} \sin^2(\pi \eta/d)
$  with Hamiltonian
\be
H_f &=&
\int  d^3r \,\bigg[\psi^\dagger (\v{r}) \lp -\frac{\hbar^2}{2m}\nabla^2 - \mu +V_p (\v{r})\rp \psi (\v{r}) \nonumber\\
    &&\hspace{0.4in}{}+  \frac{4\pi\hbar^2 a}{m} \psi^\dagger(\v{r})\psi^\dagger(\v{r})\psi(\v{r})\psi(\v{r})\bigg],\label{eq:ham-full-hom}
\ee
where $m$ is the particle mass, $\mu$ is the chemical potential, and $\psi$ and $\psi^\dagger$ are bosonic annihilation and creation operators. Adding
an additional trapping potential presents no additional difficulties.

\textit{Constructing the effective Hamiltonian.---}For each isolated site, we proceed to
build up the many-body states from the solution of the $n$-body problem at site $j$:
$\langle{\v{r_1},\cdots,\v{r_n}|}\overline{|{n}\rangle_j}= \psi_n(\v{r_1}-\v{R_j},\cdots,\v{r_n}-\v{R_j})$, which obeys
$H_j \overline{|{n}\rangle_j} =\epsilon_{n} \overline{|{n}\rangle_j} $ where  $H_j$
is the same as Eq.~\eqref{eq:ham-full-hom}'s $H_f$, except replacing the periodic potential $V_p$ there with an on-site potential $V_j$.
A convenient approximation is
to take $V_j(\v{r})=(m\omega^2/2)(\v{r}-\v{R_j})^2$ with $\omega=
2\sqrt{V_0 E_R}/\hbar$, the harmonic approximation to the site located at $\v{R_j}$.
For each site filling $n$, we
restrict our on-site basis to the lowest energy $n$-body state; however, including a finite number of excited states is straightforward.  Note that even in the non-interacting case these states  are {\em not} Wannier states.  The principle difference is that states defined in this way are non-orthogonal.
From these, however, one can construct  a new set of orthogonal states
$|{n}\rangle_j$, which hold similar physical meaning.
In the noninteracting limit, the $\ket{n}_j$ approximate the
Wannier states.

Because the single-site wavefunctions decay like Gaussians,
it typically suffices to build up the effective Hamiltonian from neighboring sites.  In particular, consider two sites $L$ and $R$, 
and the space spanned by
%for which we can build the two-site states from the tensor products 
$\overline{|n_L,n_R\rangle}=\overline{|n_L\rangle_L}\otimes\overline{ |n_R\rangle_R}$, with
overlaps
$S^{(mn)}  = \overline{\langle m,n|}\,
\overline{|{m+1,n-1}\rangle}$.
 To lowest order in the overlaps, we can define orthogonal $|n_L,n_R\rangle$ by taking
$\ket{n_L,n_R} =
\overline{\ket{n_L,n_R}}
   -
   (1/2)[ S^{(n_{R},n_{L})}\overline{\ket{n_R+1,n_L-1}}
   +S^{(n_{L},n_{R})}\overline{\ket{n_R-1,n_L+1}}
   ].$

\begin{figure}[hbtp]
\setlength{\unitlength}{1.0in}
{\includegraphics[width=3.45in,angle=0]{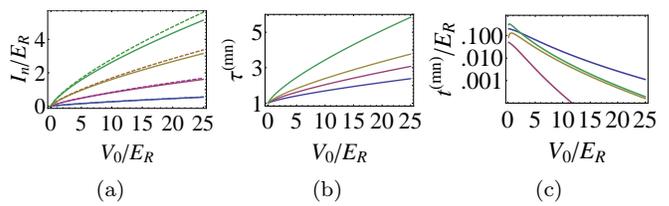}}
\caption{(color online)
(a) On-site energy with non-interacting energies subtracted off, $I_n=E_n-(3\hbar \omega/2-\mu)n$, and energies scaled by $E_R=\hbar^2 \pi^2/(2m d^2)$, using typical $^{87}$Rb parameters: lattice spacing $d=532$nm, scattering length $a=5.32$nm.  Dashed: neglecting on-site correlations, solid: including correlations.
(b) Representative hopping matrix elements with on-site correlations, relative to those neglecting on-site correlations, $\tau^{(mn)}\equiv t^{(mn)}/t\sqrt{(m+1)n}$,
as a function of lattice depth $V_0$ on a scale. Bottom to top line: $t^{(03)}, t^{(31)}, t^{(05)},t^{(35)}$.
(c) For comparison, $t^{(01)}/E_R$ calculated from the exact Wannier states  (upper curve) along with our Gaussian approximation to it with and without non-orthogonality corrections (second and third highest, respectively); also shown is the next-nearest neighbor hopping matrix element (bottom curve).
The effective Hamiltonian parameters are calculated perturbatively in $a/d$ for a Gaussian ansatz.
}
\label{fig:rb}
\end{figure}

Within this restricted basis, the effective Hamiltonian for these two sites is
$
H_{\text{eff}} =
\sum_{n,m,n',m'} \ket{n',m'}\opsandwich{n',m'}{H_f}{n,m}\bra{n,m}
$.  Evaluation to lowest order in $S^{(mn)}$ yields on-site energy terms
$\sum_{n,m} (E_n+E_m)\ket{n,m}\bra{n,m}$  and a ``hopping" term
$
-\sum_{nm} t^{(mn)}
|m+1,n-1\rangle
\langle m,n|+\hc
$
with
\be
E_n
    &=& \overline{\bra{n}}H_f\overline{\ket{n}},\nonumber\\
\!\!\!\!\! t^{(mn)} \!\!&=&\!\!
% -\langle m+1,n-1| H_f |m,n\rangle\nonumber \\&&\hspace{-0.4in}{}=
-\overline{\langle m+1,n-1| } H_f \overline{|m,n\rangle}
+ \frac{S^{(mn)}}{2}\lp E_m\!+\!E_n \rp.\label{eq:hopping-param-evaluated}
\end{eqnarray}
Additionally there is an interaction term
$U=\sum_{nm}[ U_{LL}^{(n)}+U_{RR}^{(m)}+U_{LR}^{(n,m)}]
|m,n\rangle
\langle m,n|$ with
$U_{LL}^{(m)}=U_{RR}^{(m)}=E_m$ and
\begin{eqnarray}
U_{LR}^{(n,m)}
&=&\overline{\langle m,n|} H_f \overline{|m,n\rangle}-E_m-E_n\label{eq:offsite-intern}
\end{eqnarray}
to $O(S^2)$, consistent with the rest of our calculations.
In the remainder of this paper we will neglect the off-site interaction, Eq.~\eqref{eq:offsite-intern}, and the last term in Eq.~\eqref{eq:hopping-param-evaluated}.  The former is rigorously justified as it falls of exponentially faster than the other interaction terms.  
Formally,
%The neglect of 
the non-orthogonality contribution to Eq.~\eqref{eq:hopping-param-evaluated} is 
suppressed only
 by a factor of $(V_0/E_R)^{1/4}$ with $E_R=\hbar^2\pi^2/(2md^2)$,
 but as shown in Fig.~\ref{fig:rb}, it is typically small.

The simplest many-site Hamiltonian which reduces to this one in the limit of two sites is
\be
H &=& -\sum_{\langle i,j\rangle ;m,n} t_{ij}^{(mn)}\ket{ m+1}_i\ket{n-1}_j \bra{m}_i\bra{n}_j
\nonumber\\    &&\hspace{0.3in}{}
    + \sum_{i,n}  E_n \ket{n}_i\bra{n}_i,
    \label{eq:ham-gen-hubb}
\ee
where $\sum_{\langle i,j\rangle}$ indicates a sum over nearest neighbors $i$ and $j$.
At higher order, one generates more terms such as next nearest neighbor hoppings, pair hoppings, and longer range interactions.

\textit{Calculating the Hamiltonian parameters.}---Here we consider the cases of weak interactions, resonant interactions, and coupled quantum Hall puddles.

In the limit of weak interactions, one can estimate the parameters in Eq.~\eqref{eq:ham-gen-hubb} by taking the on-site wavefunction to be 
$\psi_n\propto \exp(-\sum_{j=1}^n r_j^2/2\sigma_n^2)$,
with variational width
$\sigma_n$.  %Consequently, one can solve the on-site problem perturbatively, and the parameters in Eq.~(\ref{eq:ham-gen-hubb}) are readily calculated.
To leading order in $a/d$ we find
 $E_n =
E_R\big[(3\sqrt{V_0/E_R}-\mu/E_R)n + U n(n-1) \lp 1- \frac{3\pi}{4\sqrt{2\pi}} (a/d) (n-1)(V_0/E_R)^{1/4}\rp \big]
$
with $U=(a/d)\sqrt{2\pi}(V_0/E_R)^{3/4}$
and $t^{(mn)}=t\sqrt{n (m+1)}
[1+\frac{\sqrt{2} a \pi^{5/2} }{4d}(m+n-1)(V_0/E_R)^{3/4}]
$ with $t=V_0 (\pi^2/4-1 )e^{-(\pi^2/4)\sqrt{V_0/E_R}}$.  Note that as expected, interaction spreads out the Wannier functions, increasing the  $t^{(mn)}$'s and decreasing the $E_n$'s.
Fig.~\ref{fig:rb} shows several of the resulting $t^{(mn)}$ as a function of $V_0$ for parameters in typical optical lattice experiments with $^{87}$Rb.   Also shown are $t^{(01)}$ and the next-nearest-neighbor hopping, $t_{\text{nnn}}$, calculated from the exact Wannier states.
%Related results regarding the $n$-dependence of $t$ have  been previously obtained
Our estimates are consistent with previous work regarding $t$'s   $n$-dependence~\cite{li:moderate-filling-generalized-bose-hubbard,
mazzarella:extended-hubbard,smerzi:high-filling-on-site-Thomas-Fermi},
validating our approach.
As can be seen in Fig.~\ref{fig:rb}(c),
the relative size of the nearest neighbor hopping $t_{\text{nnn}}/t$ is $10\%$ ($1\%$) for $V_0=3E_R$ ($V_0=10E_R$),
justifying our approximation of including only nearest neighbor overlaps to describe the system near the Mott state.
\textit{We also see that
even for this weakly interacting case
the number dependence of $t$ is crucial for a quantitative description of the experiments.}  Similarly, the number dependence of the on-site interaction is quantitatively significant. This latter deficiency
of the standard Hubbard model has been noted in the past, for example by the MIT experimental group~\cite{campbell:clock-shift-MI}.

\begin{figure}[hbtp]
\centering
\hspace{-0.in}\includegraphics[width=3.50in,angle=0]{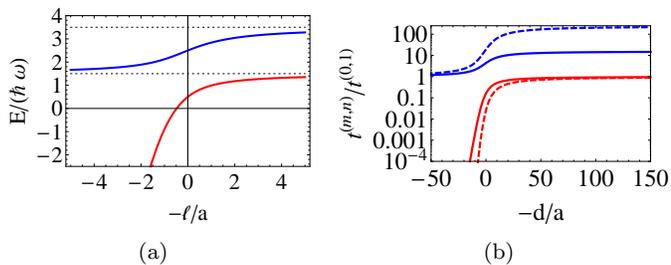}
\caption{(color online) Left: On-site two-particle energy as a function of scattering length $a$ rescaled by the on-site harmonic oscillator energy $\hbar\omega=
2\sqrt{V_0 E_R}$, for the two lowest energy branches.
  The
  corresponding characteristic length is $\ell=\sqrt{\hbar/m\omega}$.
Right: Log plot of rescaled hopping matrix elements $t^{(mn)}/(\sqrt{(m+1)n}t)$.  Solid and dashed curves are $t^{(11)}/(\sqrt{2} t)$ and $t^{(12)}/(2 t)$, respectively. We have chosen the lattice depth  $V_0=15E_R$; this affects only the horizontal scale. In the ordinary Bose-Hubbard model, $t^{(mn)}/(\sqrt{(m+1)n}t)=1$ for all $m,n$, as confirmed by this figure's $a=0^-$ lowest branch and $a=0^+$ second branch limits.  The resonance at $-d/a=0$ separates the molecular side (left), from the atomic side (right).
}
\label{fig:resonant}
\end{figure}

For more general experimental systems, one needs to include still more on-site correlations.  As our first example, we consider lattice bosons near a Feshbach resonance~\cite{daley:resonant-bosons-lattice,
dickerscheid:opt-latt-bose-feshbach}, describing, for example, ongoing Cesium atom experiments~\cite{chengchin:priv-comm}. We restrict ourselves to site occupations $n=0,1,2$, for which we have exact analytic solutions to the on-site problem for arbitrary $a$ in terms of confluent hypergeometric functions~\cite{busch:2-atoms-harmonic}.
 Fig.~\ref{fig:resonant} shows graphs of $E_n$ and $t^{(mn)}$ rescaled by $\hbar \omega$ as a function of $a$ in the deep lattice limit.

As Fig.~\ref{fig:resonant}(b) illustrates dramatically, the hopping from and to doubly-occupied sites is strongly suppressed near the Feshbach resonance when atoms occupy the lowest  branch, and is enhanced for the next-lowest branch.
The former has implications for studies of boson pairing on a lattice~\cite{daley:resonant-bosons-lattice,
dickerscheid:opt-latt-bose-feshbach}, showing that one must dramatically modify previous models near resonance,  and, as will be discussed more below, the latter implies a substantial reduction of the $n=2$ Mott lobe's size for repulsive bosons.

We give one further example, namely the case, similar to the one discussed in \cite{popp:adiabatic-make-fqh,baur:fqh-pi-pulse}, where the individual sites of the optical lattice are elliptically deformed and rotated about their center.
This is accomplished by rapidly modulating the phase of the optical lattice lasers to generate an appropriate time-averaged optical potential.  At an appropriate rotation speed $\Omega$ the lowest energy $n$-particle state on each site is a $\nu=2$ Laughlin state
$
\psi_n(\v{r_1},\ldots,\v{r_n})=\mc N_n \lb \prod_{i<j=1}^n(w_i-w_j)^2)\rb e^{-\sum_{j}\lp |w_j|^2+z_j^2 \rp/(4\ell^2)}
$,
where we define $w_j \equiv  x_j + i y_j$, and $\mc N_n$ is a normalization factor with phase chosen to gauge away phase factors appearing in $t^{(mn)}$.
  Truncating to this set of states for $n=0,1,2$, we produce an effective Hubbard model of the same form as Eq.~\eqref{eq:ham-gen-hubb}.  The hopping parameters for asymptotically deep lattices $V_0/E_R \gg 1$ are
$t^{(01)}=t$, $t^{(02)}=t(\pi^2/32) (V_0/E_R)^{1/2}$, and $t^{(12)}=t (\pi^4/1024)(V_0/E_R)$
where
$E_R = \hbar^2 \pi^2 /(2md^2)$ is the recoil energy, and $t$ is the same as in the weakly interacting case treated above;  the interaction parameters are
$E_m = \frac{\omega-\Omega}{2} m(m-1) - \mu m.$
%Naive calculations of the $t^{(mn)}$ yields phase factors, but these may be gauged away by choosing the $\mc N_n$'s phases properly.
One particularly interesting
aspect of this model of coupled quantum Hall puddles is that when the system is superfluid, the order parameter is exactly the quantity defined by Girvin and MacDonald~\cite{girvin:qh-top-order-param,read:qhe-top-order-param} to describe the nonlocal order of a fractional quantum Hall state.  Thus when one probes the superfluid phase stiffness, one directly couples to this quantity.

\textit{Mean-field theory.---}The true strength of our approach is that the resulting generalized Hubbard model
is amenable to all of the analysis used to study the standard Bose-Hubbard model.  In particular, we can gain insight from a
 Gutzwiller mean-field theory (GMFT)~\cite{jaksch:olatt,fisher:bhubb}.
 This approximation to the ordinary Bose-Hubbard model gives moderate quantitative agreement with more sophisticated methods:
 for example, the unity site filling MI/SF transition on a 3D cubic lattice occurs at $(t/U)_c=0.034 08(2)$ while GMFT yields $(t/U)_c = 0.029$~\cite{capogrosso-sansone:mi}.

In the ground state $\ket{\Psi}$, we introduce mean fields
$\xi_m
\equiv \overlap{\Psi}{m+1}\overlap{m}{\Psi}
$.
Neglecting terms which are quadratic in $\delta L_m^i=\ket{i,m+1}\bra{i,m}-\xi_m$,
the Hamiltonian is
$
H_{MF} = \sum_i H_{MF,i}
$
with
\be
H_{MF,i}
     &=& E_{n_i} \ket{n}_i\bra{n}_i
       \nonumber\\&&\hspace{-0.525in}{}
     -z\sum_m \bigg[ \zeta_m \ket{m-1}_i\bra{m}_i
    - \zeta_m \xi_{m-1}  +\hc  \bigg],\label{eq:on-site-HMF-final}
\ee
where $\hc$ denotes Hermitian conjugate, $z$ is the lattice coordination number, and
$\zeta_n = \sum_m \xi_m t^{(mn)} $.

\begin{figure}[!t]
\centering
\includegraphics[width=3.00in,angle=0]{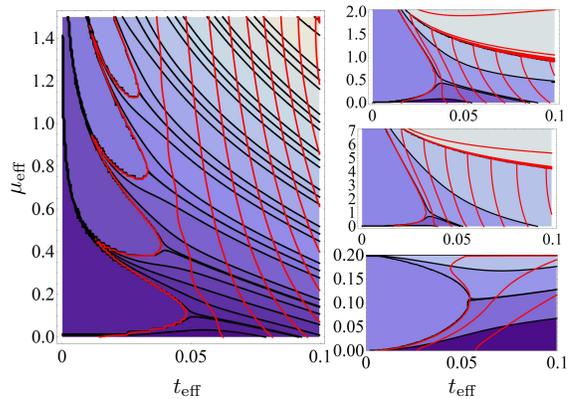}
\caption{(color online)
Representative Gutzwiller mean-field theory phase diagrams, showing constant density (black, roughly horizontal) and constant $\xi\equiv\zeta_1+\zeta_2+\zeta_3$ (red, roughly vertical) contours: $\xi $, similar to the condensate density, is a  combination of the mean fields $\zeta_m$, defined after Eq.~(\ref{eq:on-site-HMF-final}).  Density contours are   $n=\{0.01, 0.2, 0.5, 0.8, 0.99, 1.01, 1.2, \ldots\}$ and order parameter contours are $\xi=\{0.2,0.4,\ldots\}$, except (d) where we take contours  $\xi=\{0.02,0.04,\ldots\}$.
The phase diagrams are  functions of $\mu_{\text{eff}}\equiv \mu/E_{\text{rec}}$ and $t_{\text{eff}}\equiv \exp \lp - \sqrt{V_0/E_{\text{rec}}}\rp$, where the lattice depth $V_0$ is the natural experimental control parameter. We plot versus $t_{\text{eff}}$, instead of $V_0$, as this is closer to the Hamiltonian matrix elements and  more analogous to traditional visualizations of the Bose-Hubbard phase diagram.
(a) Ordinary Bose-Hubbard model for $a=0.01d$, (b)  resonant lattice bosons restricted to fillings $n=0,1,2$  with $a=0.01d$, on the next to lowest energy branch on the $a>0$ side of resonance, (c) resonant lattice boson model with $a=d$, and (d) FQH puddle array model taking $\omega-\Omega=0.1E_{\text{rec}}$ (see text for details).
}
\label{fig:phase-diagram-representative-cases}
\end{figure}

Truncating the number of atoms on a site to $n\leq n_{\rm max}$, we self-consistently solve Eq.~(\ref{eq:on-site-HMF-final}) by an iterative
method.  We
start with trial mean-fields, calculate the lowest energy eigenvector of the $(n_{\rm max}+1)\times (n_{\rm max}+1)$ mean field Hamiltonian matrix, then update the mean-fields.   We find that it typically suffices to take $n_{\rm max}$ roughly three times the mean occupation of the sites.
Fig.~\ref{fig:phase-diagram-representative-cases} illustrates how the
density dependence of the parameters introduced by the on-site correlations modify the GMFT phase diagram --- particularly the phase boundary's shape, and the density and order parameter in the superfluid phase.

\begin{figure}[!t]
\centering
\includegraphics[width=3.50in,angle=0]{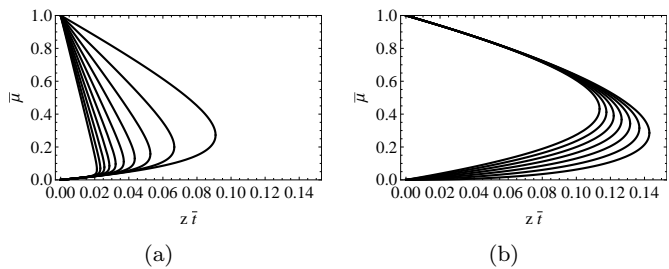}
\caption{Single lobe of the Mott insulator/superfluid boundary.
Complete characterization of the mean-field lobe shape  in the $z \bar t$-$\bar \mu$ plane for all possible $t^\pm$'s (see Eq.~\eqref{eq:resc-vars} for definitions).
(a) Fix $t^-=0.5$, vary $t^+$ from 1 (outer curve) to 21 (inner curve) in steps of 2.   (b) Fix $t^+=1.5$, vary $t^-$ from 0 (outer curve) to 1 (inner curve) in steps of $0.15$.
\label{fig:phase-diagram-1-lobe-fully-characterized}
}
\end{figure}

  As one would expect,
  the topology of the MI/SF phase boundaries
  are similar to that of the standard Bose-Hubbard model, but the
Mott
lobes' shapes can be significantly distorted.
Within mean-field theory the
boundary's shape can be determined analytically by taking
 $\ket{\Psi}=\epsilon' \ket{n-1}+\sqrt{1-\epsilon'^2-\epsilon^2}f_{n-1}\ket{n} + \epsilon \ket{n+1}$,
 and expanding $\opsandwich{\Psi}{H_{MF}}{\Psi}$ to quadratic order in $\epsilon$ and $\epsilon^\prime$.  The Mott boundary corresponds to when the energy expectation value's Hessian changes sign; this boundary occurs when
\be
\lp
E_{n+1}-  E_{n}
+ 2z t^{(n,n+1)}\rp  &&\nonumber\\
&&\hspace{-1.6in}{}\times
\lp
E_{n-1} -  E_n
+ 2z t^{(n-1,n)}\rp=  \lp 2 z t^{(n,n)}\rp^2.\label{eq:analytic-mott-lobe}
\ee
The  five scaled parameters
\be
&&\hspace{-0.1in}\bar\mu \equiv \frac{E_{n}-E_{n-1}}{E_n},
\hspace{0.25in} x_U \equiv \frac{E_{n+1}+E_{n-1} -2E_n}{E_n},\nonumber\\
&& \bar t \equiv \frac{t^{(n,n)}}{E_n}, \hspace{0.25in}
  t^+ \equiv \frac{t^{(n,n+1)}}{t^{(n,n)}},
    \hspace{0.25in}
    t^- \equiv \frac{t^{(n-1,n)}}{t^{(n,n)}},
  \label{eq:resc-vars}
   \ee
 \textit{completely} characterize the shape of a filling-$n$ Mott phase boundary.
   Varying $\bar t$ and $\bar\mu$ while fixing the other parameters then maps out a Mott-lobe like feature in the $\bar t$ and $\bar\mu$ plane, is illustrated in Fig.~\ref{fig:phase-diagram-1-lobe-fully-characterized}.

\textit{Summary and discussion.---}We have demonstrated a novel approach to strongly correlated lattice boson problems.
One constructs a model by truncating the on-site Hilbert space to a single state for  each site filling $n$ and includes only nearest-site, single particle hoppings.  While this approximation captures many multi-band effects, it is not a multi-band Hubbard model and in particular retains the ordinary Bose-Hubbard model's simplicity. In this method, arbitrary on-site correlations may be treated, even in the mean field theory, and consequently it
captures the condensate depletion, modified excitation spectra, altered condensate wavefunction, and altered equation of state characteristic of strongly interacting bosons.
We have calculated the mean field Mott insulator/superfluid phase boundary analytically and observables across the phase diagram numerically.

Finally, although we have
truncated to a single many-body state for each filling $n$,
no difficulty arises from
including on-site many-body excitations in the Hamiltonian. These are especially important, for example, for double well lattices and spinor bosons.
The ideas also extend straightforwardly to fermions; see Refs.~\cite{diener:fermion-opt-lattice-feshbach,zhai:SF-ins-fermi-opt-latt,koehl:lattice-fermions-feshbach,mathy:neel-temp-not-1band} for related considerations.

\textit{Acknowledgments.}---This material is based upon work supported by the
National Science Foundation through grant No. PHY-0758104. We
thank Stefan Baur and Mukund Vengalattore for useful conversations.

%\bibliography{fqh-puddles-w-hopping}

\begin{thebibliography}{10}

\bibitem{bloch:many-body-cold-atoms-review}
I. Bloch, J. Dalibard, and W. Zwerger, Reviews of Modern Physics (to appear)
  (2008).

\bibitem{jaksch:olatt}
D. Jaksch {\it et~al.}, Phys. Rev. Lett. {\bf 81},  3108  (1998).

\bibitem{chengchin:priv-comm}
Cheng Chin, private comm.

\bibitem{popp:adiabatic-make-fqh}
M. Popp, B. Paredes, and J.~I. Cirac, Phys. Rev. A {\bf 70},  053612  (2004).

\bibitem{baur:fqh-pi-pulse}
S.~K. Baur, K.~R.~A. Hazzard, and E.~J. Mueller, Physical Review A (Atomic,
  Molecular, and Optical Physics) {\bf 78},  061608  (2008).

\bibitem{barmettler:double-wells}
P. Barmettler {\it et~al.}, Physical Review A (Atomic, Molecular, and Optical
  Physics) {\bf 78},  012330  (2008).

\bibitem{jiang:double-well-decoherence-free}
L. Jiang {\it et~al.}, Physical Review A (Atomic, Molecular, and Optical
  Physics) {\bf 79},  022309  (2009).

\bibitem{li:moderate-filling-generalized-bose-hubbard}
J. Li, Y. Yu, A.~M. Dudarev, and Q. Niu, New Journal of Physics {\bf 8},  154
  (2006).

\bibitem{mazzarella:extended-hubbard}
G. Mazzarella, S.~M. Giampaolo, and F. Illuminati, Phys. Rev. A {\bf 73},
  013625  (2006).

\bibitem{smerzi:high-filling-on-site-Thomas-Fermi}
A. Smerzi and A. Trombettoni, Phys. Rev. A {\bf 68},  023613  (2003).

\bibitem{campbell:clock-shift-MI}
G.~K. Campbell {\it et~al.}, Science {\bf 313},  649  (2006).

\bibitem{daley:resonant-bosons-lattice}
A.~J. Daley {\it et~al.}, arxiv {\bf 0810.5153},    (2008).

\bibitem{dickerscheid:opt-latt-bose-feshbach}
D.~B.~M. Dickerscheid, U.~A. Khawaja, D. van Oosten, and H.~T.~C. Stoof, Phys.
  Rev. A {\bf 71},  043604  (2005).

\bibitem{busch:2-atoms-harmonic}
T. Busch, B.-G. Englert, K. Rzazewski, and M. Wilkens, Foundations of Physics
  {\bf 28},  549  (1998).

\bibitem{girvin:qh-top-order-param}
S.~M. Girvin and A.~H. MacDonald, Phys. Rev. Lett. {\bf 58},  1252  (1987).

\bibitem{read:qhe-top-order-param}
N. Read, Phys. Rev. Lett. {\bf 62},  86  (1989).

\bibitem{fisher:bhubb}
M.~P.~A. Fisher, P.~B. Weichman, G. Grinstein, and D.~S. Fisher, Phys. Rev. B
  {\bf 40},  546  (1989).

\bibitem{capogrosso-sansone:mi}
B. Capogrosso-Sansone, N.~V. Prokof'ev, and B.~V. Svistunov, Phys. Rev. B {\bf
  75},  134302  (2007).

\bibitem{diener:fermion-opt-lattice-feshbach}
R.~B. Diener and T.-L. Ho, Phys. Rev. Lett. {\bf 96},  010402  (2006).

\bibitem{zhai:SF-ins-fermi-opt-latt}
H. Zhai and T.-L. Ho, Phys. Rev. Lett. {\bf 99},  100402  (2007).

\bibitem{koehl:lattice-fermions-feshbach}
M. K\"{o}hl {\it et~al.}, Phys. Rev. Lett. {\bf 94},  080403  (2005).

\bibitem{mathy:neel-temp-not-1band}
C. Mathy and D.~A. Huse,  {\bf arxiv:0805.1507},    .

\end{thebibliography}

\end{document}